\documentclass[a4paper,12pt]{article}

\newcommand{\beqa}{\begin{eqnarray}}
\newcommand{\eeqa}{\end{eqnarray}}
\newcommand{\flux}{{\rm cm^{-2}s^{-1}sr^{-1}}}
\usepackage{amsmath} 
\usepackage{epsfig}

\begin{document}
\begin{titlepage}
\title{Neutrino clustering and the Z-burst model}
\author{B.\ H.\ J.\ McKellar\thanks{e-mail b.mckellar@physics.unimelb.edu.au} 
\, and\, M.\ Garbutt\thanks{e-mail mag@physics.unimelb.edu.au} \\
University of Melbourne \\
Parkville, Victoria 3052, Australia\\
\\
G.\ J.\ Stephenson Jr.\ \thanks{e-mail: gjs@baryon.phys.unm.edu}\\
University of New Mexico \\
Albuquerque, New Mexico 87131 \\
\\
T.\ Goldman\thanks{e-mail goldman@t5.lanl.gov} \\
Los Alamos National Laboratory \\
Los Alamos, New Mexico 87545 \\
}

\date{June 04, 2001}
\maketitle
\vspace*{-5.1in}
\flushright{LA-UR-01-3234}
\vspace{-.1in}
\flushright{UM-P-2001/016}
\vspace{-.4in}
\vspace*{4.9in}

\begin{abstract}

The possibility that the observed Ultra High Energy Cosmic
Rays are generated by high energy neutrinos creating
``Z-bursts" in resonant interactions with the background
neutrinos has been proposed, but there are difficulties in
generating enough events with reasonable incident neutrino
fluxes.

We point out that this difficulty is overcome if the
background neutrinos have coalesced into ``neutrino clouds"
--- a possibility previously suggested by some of us in
another context.  The limitations that this mechanism for
the generation of UHECRs places on the high energy neutrino
flux, on the masses of the background neutrinos and the
characteristics of the neutrino clouds are spelled out.
\end{abstract}
\end{titlepage}

\section{Introduction}

The continuation of the cosmic ray spectrum beyond the GZK (Greisen,
Zatsepen and Kuzmin) cutoff is a puzzle which has inspired many
attempts at explanation.  Neutrino-anti-neutrino annihilation into a
Z-boson which subsequently decays to hadrons and leptons is one
promising solution of the problem.  Should relic neutrinos possess a
mass in the $\rm{eV/c^2}$ range the center of momentum energy required for
resonant annihilation can be produced by cosmic ray neutrinos of
energy $\sim 10^{21}\ \rm{eV}$.  This proposal was first examined in
detail by Weiler \cite{weiler3} \cite{weiler1} and Fargion
\cite{Fargion:1999ft} , who assumed that the relic neutrinos cluster
in galaxies up to two or three times the mean relic density, and found
that the required flux of cosmic ray neutrinos is larger than had
previously been suggested.  It has been suggested that the clustering 
may extend to densities of $100$ to $10 000$ times the normal relic 
density on galactic scales, to moderate the required incident 
flux \cite{B-P}.

Independently, it has been shown that the relic neutrinos may have a local
density, on scales of the solar system to a few parsecs, 
which may be many orders of magnitude higher than
standard cosmology predicts without conflicting with any known
experimental results \cite{Mckellar1}.  Furthermore, 
the formation mechanism provides a strong argument for a natural
association between the so called neutrino clouds of
Ref.~(\cite{Mckellar1}) with  
small celestial objects such as stars and solar systems.

In this note we investigate the consequences of this latter scenario for
the measured Z-burst rate and for the bounds on the neutrino cloud. 
Specifically, we find the neutrino cloud radius and density range
required to reproduce the observed flux of UHECRs and use the
non-observation of coincident UHECR events to strengthen further these
constraints.

We note that recent studies \cite{Pas:2001nd, Fodor:2001rg} of the Z-burst
explanation for UHECRs have derived information about the masses of
the relic neutrinos.  In the light of the present study of a dense 
local cloud of neutrinos as the target, we note that the Fermi motion 
of the neutrinos in the local cloud would complicate attempts to 
determine the neutrino mass spectrum from the observations of UHECRs.



\section{Numbers and formalism}
\subsection{Z-burst flux}\label{zFlux}
To evaluate the Z-burst flux at earth we will follow closely the
formalism used by Weiler \cite{weiler1}.  The change in the
differential neutrino
flux density $F_\nu(E_\nu,x)$ at a given energy $E_\nu$ and a distance $x$
from its source is 
given by
\beqa
dF_\nu (E_\nu,x) = -F_\nu (E_\nu,x)\sigma_\nu (E_\nu)n_\nu(x) dx,
\eeqa
where $\sigma_\nu (E_\nu)$ is the annihilation cross section, and
$n_\nu(x)$ is the local neutrino density.  The reduction in neutrino flux
found at a distance  $D$ from the source is therefore found to be
\beqa
\delta F_\nu (D)&=& \int dE_\nu F_\nu(E_\nu ,0) 
\biggl(1-\exp\{-\sigma(E_\nu )S(D)\}\biggr),
\eeqa
where $S(D)=\int^D_0 n_\nu (x)dx$ is the column density, and $\delta
F_\nu (D) = F_\nu (0)-F_\nu (D)$.

In the narrow resonance approximation, where the variation in
$F_\nu(E_\nu ,0)$ over the resonant energy is small, we can write
\beqa
\delta F_\nu (D) &=&  E_R F_\nu(E_R,0)\int {ds\over M_Z^2}
\biggl(1-\exp\{-\sigma(s)S(D)\}\biggr), 
\eeqa
where the resonance energy $E_R$ corresponds to a neutrino energy such
 that the center of momentum energy is $s=M_Z^2$.  From this point on
 references to the neutrino flux should be taken to mean the quantity
 $E_R F_\nu(E_R,0)$ with units $\flux$. 
 The main contribution to $\sigma(s)$ is given
by the s-channel Z-resonance, $\nu\bar{\nu}\to f\bar{f}$, with a
 branching ratio of 70\% to hadrons, 20\% to neutrinos and 10\% to
 charged leptons\cite{roulet},
\beqa
\sigma_Z(s)&=&\sum_f {2G_F^2\over
3\pi}n_fsM_Z^4\biggl[{t^2_3(f)-2t_3(f)Q_fs^2_W +2Q_f^2s_W^4\over
(s-M_Z^2)^2+ \Gamma_Z^2M_Z^2}\biggr]
\eeqa
where the quantities in the square brackets are the standard model
weak isospin numbers, charge and sine of the Weak angle.  In our
calculation we set $\sigma(s)=\sigma_Z(s)$.

The only quantity entering these considerations which has been
measured directly are the post GZK events, with fluxes quoted as
$F_{p/\gamma}= 10^{-19}\rm{\flux} $ above $5\times 10^{19}\ \rm{eV}$
and $F_{p/\gamma}\sim 2\times 10^{-20}\rm{\flux} $ above $5\times
10^{20}\ \rm{eV}$ \cite{weiler1}\cite{Nagano:2000ve}.  Assuming the
Z-burst flux equals $\delta F_\nu (D)$, and that the Z's are then
converted to nucleons and photons, we can write
\beqa 
F_{p/\gamma}=M\times \delta F_\nu (D) , 
\eeqa
where $M$ is the photon and nucleon multiplicity per Z-burst, estimated
to be $\sim 30$.  The aim of any calculation must be to reproduce the
measured value of 
$F_{p/\gamma}$.  

\subsection{Neutrino flux}\label{nuFlux}
The Z-burst model, be it in the presence of a non-clustered relic
neutrino background or a neutrino cloud, requires a flux of UHE neutrinos.
It is no surprise that the neutrino flux has not been 
directly measured  at UHE energies; however there are theoretical as well as
experimental upper bounds.

The experimental upper bounds are derived in Ref.~(\cite{weiler1}) from 
the non-observation of deeply penetrating particles with energy
$>10^{17}\ \rm{eV}$ at the Fly's Eye detector in Utah \cite{balt}.
The upper bounds on the flux are: $2\times 10^{-13}$, $3\times
10^{-14}$ and $5\times 10^{-16}\ \rm{cm^{-2}s^{-1}sr^{-1}}$ for $E_\nu=
10^{17}$, $10^{18}$ and $10^{20}\ \rm{eV}$ respectively.  This is
unfortunately not able to constrain the Z-burst model.  In the
absence of experimental data one needs to turn to models of
astrophysical particle acceleration.


The mechanism by which primaries with energies above
$10^{19}\ \rm{eV}$ are produced
 is as much a puzzle as their propagation to the earth. Neutrino production
resulting from the interaction of nucleons produced in AGN
jets or gamma-ray bursts interacting with the cosmic microwave
background is one explanation; other explanations involve
more exotic physics such as decaying super heavy particles, or
topological defects. For a review of such theories and a discussion of
the expected neutrino flux see Ref.~(\cite{Bhattacharjee:2000qc}) and
references therein.  The former explanation is the
only theoretical 
description which is non-speculative in that it uses only standard 
model physics.  There is however an ongoing debate as to the magnitude
of the neutrino flux which may be derived from conventional 
sources \cite{waxman}.  Depending on the assumptions made the
upper-bound to the neutrino flux from astrophysical sources ranges from
$E_RF_\nu(E_R,0) \leq 3\times 10^{-20}$ to $1\times 10^{-19}\flux$.

\subsection{Cloud density and radius}
While the neutrino cloud can be formed with a complex density profile,
e.g. with more massive species at the center \cite{Mckellar1}, we
parameterise the cloud as a sphere with no diffuse boundary such that
the density of the relic neutrinos is
\beqa
n_\nu(x) = n_{c}(x)\theta (R-x) + n_{0}(x)\theta (x-R),
\eeqa 
where $n_{c}(x)$ is the density inside the cloud and $n_{0}(x)$ is the
density outside.  In the standard theory the density of relic
neutrinos is $n_0(x)= 54{\rm cm}^{-3}$; the formation of neutrino
clouds would deplete this number to a negligible quantity, such that in our
calculations it can be safely ignored.    
A more complex model of the cloud structure would include a sum
over the different mass eigenstates of the neutrino species, as well
as a diffuse boundary, and perhaps some directional and temporal
dependence taking into account cloud dynamics.  However these
additions will not result in large changes to the calculated flux.  We
consider relic and cosmic 
ray neutrinos with a mass of $1\rm{eV/c^2}$, and clouds with radii in
the range of $10^{14}-10^{20}\ \rm{cm}$ \footnote{In more familiar
units $\sim 7-7\times
10^{6}\ \rm{au} $ or $\sim 3\times 10^{-5}-30\ \rm{pc}$} and density in
the range of $10^{10}-10^{16}\ \rm{cm^{-3}}$.  These parameters are
not in conflict with the requirement that the total mass of dark matter
within the orbit of Uranus be less than $\sim 3\times10^{-6}M_\odot$ \cite{anderson}.

\section{Results for neutrino clouds}


Clouds of neutrinos with density ranging from $10^{10}\ \rm{cm^{-3}}$
to $10^{16}\ \rm{cm^{-3}}$ have a correspondingly large range in Fermi
energy.  In fact $1\ \rm{eV/c^2}$ mass neutrinos in a $10^{10}\
\rm{cm^{-3}}$ cloud are non-relativistic, while in a $10^{16}\
\rm{cm^{-3}}$ cloud they are very much relativistic.  In the following
analysis we temporarily neglect the effects of the Fermi motion for 
clarity in the exposition, returning to it at a later point.

\begin{figure}
\begin{picture}(0,0)%
\includegraphics{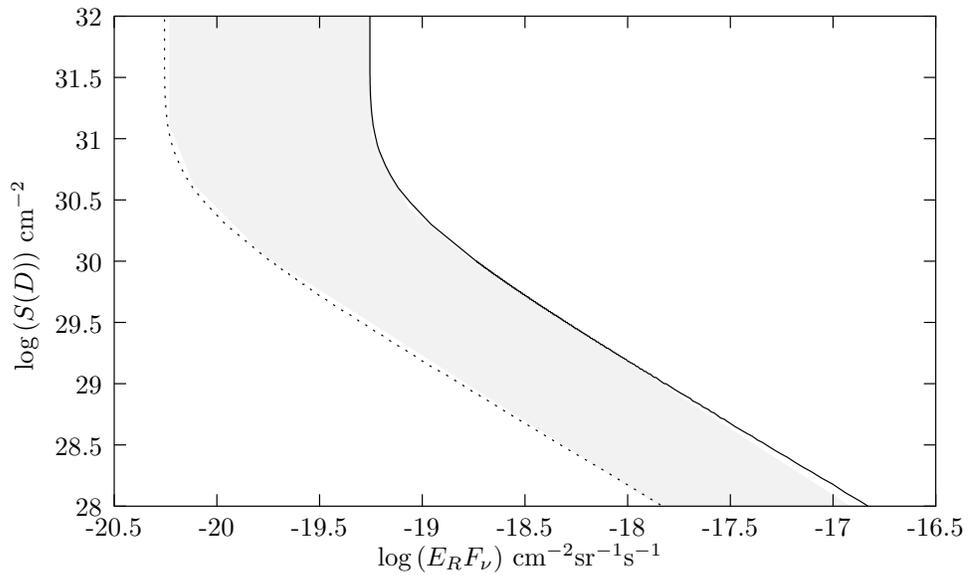}%
\end{picture}%
\setlength{\unitlength}{3947sp}%
\begingroup\makeatletter\ifx\SetFigFont\undefined%
\gdef\SetFigFont#1#2#3#4#5{%
  \reset@font\fontsize{#1}{#2pt}%
  \fontfamily{#3}\fontseries{#4}\fontshape{#5}%
  \selectfont}%
\fi\endgroup%
\begin{picture}(5798,3495)(305,-3061)
\put(857,-2751){\makebox(0,0)[rb]{\smash{\SetFigFont{10}{12.0}{\familydefault}{\mddefault}{\updefault}28}}}
\put(857,-2366){\makebox(0,0)[rb]{\smash{\SetFigFont{10}{12.0}{\familydefault}{\mddefault}{\updefault}28.5}}}
\put(857,-1981){\makebox(0,0)[rb]{\smash{\SetFigFont{10}{12.0}{\familydefault}{\mddefault}{\updefault}29}}}
\put(857,-1596){\makebox(0,0)[rb]{\smash{\SetFigFont{10}{12.0}{\familydefault}{\mddefault}{\updefault}29.5}}}
\put(857,-1211){\makebox(0,0)[rb]{\smash{\SetFigFont{10}{12.0}{\familydefault}{\mddefault}{\updefault}30}}}
\put(857,-826){\makebox(0,0)[rb]{\smash{\SetFigFont{10}{12.0}{\familydefault}{\mddefault}{\updefault}30.5}}}
\put(857,-441){\makebox(0,0)[rb]{\smash{\SetFigFont{10}{12.0}{\familydefault}{\mddefault}{\updefault}31}}}
\put(857,-56){\makebox(0,0)[rb]{\smash{\SetFigFont{10}{12.0}{\familydefault}{\mddefault}{\updefault}31.5}}}
\put(857,329){\makebox(0,0)[rb]{\smash{\SetFigFont{10}{12.0}{\familydefault}{\mddefault}{\updefault}32}}}
\put(931,-2875){\makebox(0,0)[b]{\smash{\SetFigFont{10}{12.0}{\familydefault}{\mddefault}{\updefault}-20.5}}}
\put(1576,-2875){\makebox(0,0)[b]{\smash{\SetFigFont{10}{12.0}{\familydefault}{\mddefault}{\updefault}-20}}}
\put(2221,-2875){\makebox(0,0)[b]{\smash{\SetFigFont{10}{12.0}{\familydefault}{\mddefault}{\updefault}-19.5}}}
\put(2866,-2875){\makebox(0,0)[b]{\smash{\SetFigFont{10}{12.0}{\familydefault}{\mddefault}{\updefault}-19}}}
\put(3511,-2875){\makebox(0,0)[b]{\smash{\SetFigFont{10}{12.0}{\familydefault}{\mddefault}{\updefault}-18.5}}}
\put(4156,-2875){\makebox(0,0)[b]{\smash{\SetFigFont{10}{12.0}{\familydefault}{\mddefault}{\updefault}-18}}}
\put(4801,-2875){\makebox(0,0)[b]{\smash{\SetFigFont{10}{12.0}{\familydefault}{\mddefault}{\updefault}-17.5}}}
\put(5446,-2875){\makebox(0,0)[b]{\smash{\SetFigFont{10}{12.0}{\familydefault}{\mddefault}{\updefault}-17}}}
\put(6091,-2875){\makebox(0,0)[b]{\smash{\SetFigFont{10}{12.0}{\familydefault}{\mddefault}{\updefault}-16.5}}}
\put(425,-1211){\rotatebox{90.0}{\makebox(0,0)[b]{\smash{\SetFigFont{10}{12.0}{\familydefault}{\mddefault}{\updefault}$\log{(S(D))}\ \rm{cm^{-2}}$ }}}}
\put(3511,-3061){\makebox(0,0)[b]{\smash{\SetFigFont{10}{12.0}{\familydefault}{\mddefault}{\updefault}$\log{(E_RF_\nu)}\ \rm{cm^{-2}sr^{-1}s^{-1}}$ }}}
\end{picture}
\caption{The dashed line corresponds to the required column density
needed to produce a photon/hadron flux of $10^{-20}\flux$, while the
solid line is the column density need for a photon/hadron flux of
$10^{-19}\flux $.  The shaded region corresponds to the experimentally
allowed column density.} 
\label{column}
\end{figure}

The crucial element in our analysis is the column density $S(D)$. In
Fig.~\ref{column} the column density needed to produce $F_{p/\gamma}$
for various values of neutrino flux is shown.  The upper curve
corresponds to a flux of $F_{p/\gamma}=10^{-19}\ \rm{\flux}$ while the lower curve
corresponds to $F_{p/\gamma}=10^{-20}\ \rm{\flux}$. We would expect
the actual
value lies between these two values.  
The minimum column density for a source of neutrinos traveling $50\
\rm{Mpc}$ to earth, assuming a flat distribution of relic neutrinos with a
density of $54\ \rm{cm^{-3}}$, is   
approximately $\rm{10^{28}cm^{-2}}$,
hence the column density is not evaluated below the 
point where the cosmic
rays produced on the most dilute reasonable background exceeds the
observed flux above the GZK cut off. 
With the clustering of relic neutrinos on scales described in
Ref.~(\cite{weiler1})  the column density is increased to $S(D)\sim 10^{29}\ \rm{cm^{-2}}$, while for 
a neutrino cloud with radius $R=50\ \rm{pc}$ and density
$n(x) = 10^{14}\ \rm{cm^{-3}}$ we have $S(D)\sim 10^{34}\
\rm{cm^{-2}}$.  This makes the demands on the incident neutrino flux
much more modest.

Using the fact that the flux of UHECRs is $10^{-20}\leq
F_{p/\gamma}\leq 10^{-19}$ we can establish an allowed neutrino cloud
parameter range, depending of course on the incident neutrino flux.
In Sect.~\ref{nuFlux}. we saw that the magnitude of the cosmic neutrino flux is
open to question.  Most models place an upper bound of $E_RF(E_R,0)\leq
10^{-19}\flux$,  while the lowest expected flux is not at all known.
However in the context of the Z-burst model we can define a minimum
required flux, that is, the smallest neutrino flux able to produce
$F_{p/\gamma}$ assuming the incident neutrinos are absorbed in their
entirety.  
\beqa
E_R F_{\nu\ min}(E_R,x) &=& {M_Z F_{p/\gamma}\over M\Gamma_Z}\ .
\eeqa  
This situation would occur if the average interaction length of the
neutrino becomes significantly less than the cloud radius.  Depending
on the value of $F_{p/\gamma}$, $10^{-20}\flux$ or $10^{-19}\flux$ the
corresponding minimum neutrino flux is $5.6\times 
10^{-21}$ or $5.6\times 10^{-20}\flux $ respectively.  
The minimum
neutrino flux is approached for 
high column density and small neutrino flux in Fig.~1.


The neutrino cloud radius and density permitted by the
observed UHE CR data are plotted in Figs.~\ref{large}~and~\ref{small}.  The
neutrino flux in Fig.~\ref{large} is that of the upper bound
$E_RF_\nu(E_R,0)=10^{-19}\flux$, and for Fig.~\ref{small}, just above
the minimum required, $E_RF_\nu(E_R,0)=10^{-20}\flux$.  In the first
case the observed UHECRs can be fully accounted for, while in the
second a maximum flux of $F_{p/\gamma}\sim 1.7\times 10^{-20}\flux$ is
produced.

We now return to the question of the Fermi motion of the neutrinos in
the cloud.  This motion will alter the value of the resonant energy
$E_R$.  For a cosmic neutrino interacting with a relic neutrino
$s = 2(E_\nu E_c - \vec{p}_\nu\cdot\vec{p}_c)$ where $E_c$ and $\vec{p}_c$
are the energy and momentum of the relic neutrino, at resonance and
for a relic neutrino at rest $2m_\nu E_\nu = M_Z^2$, hence $E_R={M_Z^2
\over 2m_\nu}$.  In a neutrino cloud of high density $E_R$ will take
on a range of values from $E_R={M_Z^2 \over 2m_\nu}$ to $E_R = {M_Z^2
\over 2}(E_f - p_f \cos\theta)^{-1}$, where $E_f$ and $p_f$ are the
Fermi energy and momentum of the cloud.  To account for the smearing
of resonant energy we define an average resonant energy $\bar{E}_R$,
\beqa \bar{E}_R & = & {1\over 2N}\int^{p_f}_0 {\rm d}^3p_c{M_Z^2\over
(E_c - p_c \cos\theta)}\nonumber\\
 &=& 4\pi{M_Z^2 \over 2N}\int^{p_f}_0 {\rm d}p_c p_c{\rm sinh}^{-1}\left({p_c\over
 m_\nu}\right) \ ,
\eeqa
where we have normalized to the  Fermi momentum of the cloud, $N={4\pi\over
3}p_f^3$.  The resulting average energy is
\beqa
\bar{E}_R & = & 4\pi{M_Z^2 \over 8N}\left[\left(2p_f^2+m_\nu^2\right)
{\rm sinh}^{-1}\left({p_f\over m_\nu}\right) - p_f\sqrt{p_f^2 + m_\nu^2}\right],
\eeqa
which in the non-relativistic limit, as expected, reduces to
\beqa
\bar{E}_R & = & 4\pi{M_Z^2 p_f^3 \over 6N}={M_Z^2\over 2m_\nu}\ .
\eeqa
In the case of a very dense cloud, with a Fermi momentum of $p_f \simeq 13\
\rm{eV/c}$ and for $1{\rm eV/c^2}$ mass neutrinos,
the resultant average resonant energy is $\bar{E}_R =
1.3\times 10^{21}\ {\rm eV}$.  Since our calculations have been performed in
the center 
of momentum frame the reduction in resonant energy will not significantly
alter any of our conclusions regarding the results discussed so far,
in as much as the narrow resonance approximation holds over the width
of the average energy distribution.  


As an aside we note that
the lowering of the resonant energy also allows the Z-burst
mechanism to remain viable should neutrinos possess very small
masses.  In a non-clustered background the resonant energy for a $0.1\
\rm{eV/c^2}$ mass neutrino, (rather than a $1\
\rm{eV/c^2}$ neutrino) is $E_R = 4\times 10^{22}\ \rm{eV}$, while in a
dense neutrino cloud the average resonant energy can be as low as
$\bar{E}_R=2.3\times 10^{21}\ \rm{eV}$ making the Z-burst mechanism 
viable even if the differential neutrino flux drops off very rapidly 
with energy.

The standard deviation of the resonant energy, both in absolute terms 
and relative to the mean energy, increases with increasing Fermi energy.
\begin{equation}
    (\Delta E)^{2} = (E_{R})_{0}^{2} - (\bar{E}_R)^{2}.
\end{equation}
However, for our present estimates we simply work with the mean 
resonant energy, and ignore the variation in flux over the width of the 
distribution.

A stronger constraint on the parameter range of the neutrino
cloud comes from the non-observation of coincident CR events at
ultra-high energies.  The decay products of the Z-resonance are
boosted into a cone of angle $10^{-11}$, thus occupying an opening area of
$A_p=\pi(\tan[10^{-11}]R_c)^2$, where $R_c$ is the radius at which the
neutrino annihilation took place.  If the detection area of an
experiment ($A_{\rm det}$) is larger than $A_{\rm p}/M$ then, on average a
coincident event would be observed.  The multiplicity in this context
should count all particles produced at resonance not just those
with an energy above the GZK cutoff.

With a multiplicity of $\sim 60$ and a detection area of $A_{\rm
det}\sim 500 {\rm km^2}$ a sphere of radius $R_c\sim 10^{18}{\rm cm}$
can be defined within which neutrino annihilations will on average
result in the observation of a coincident event.  The flux of
coincident events ($F_{\rm c}$)  will depend on the neutrino flux incident upon this
sphere, and can be evaluated in the formalism of Sect.~\ref{zFlux}.
\beqa
F_{\rm c} &=&  \delta F(E_\nu,x\geq
(D-R_c))\nonumber\\
&=& \biggl(\delta F(E_\nu, D) -\delta F(E_\nu,
D-R_c)\biggr)\nonumber\\
&=& M\times E_R F(E_R, 0)\int{ds\over M_Z^2}\biggl(
-\exp[-\sigma(s)S(D)]\nonumber\\
&+& \exp[-\sigma(s)S(D-R_c)]\biggr). 
\eeqa  
If in addition we make the simplifying assumption that the relic
neutrino column density outside the cloud is much less than the column
density inside we are able to set $n(x)=n_c(x)$. Now the expression for the
flux of coincident events becomes
\beqa
F_{\rm c}&=&E_R F(E_R, 0)\int{ds\over M_Z^2}
\exp[-\sigma(s)n_c R]\nonumber\\
&\times&\biggl(\exp[\sigma(s)n_c R_c]-1\biggr).
\eeqa

The non-observation of a coincident flux implies the following bound
on the total number $(N_t)$ of such events observed by a particular experiment
\beqa
 N_t={\cal A}\ t_r F_{\rm c}< 1,
\eeqa
where  ${\cal A}$ is the experimental aperture at $E_0>
10^{19.6}{\rm eV}$ and $t_r$ is the experimental running time.
In order for the neutrino cloud hypothesis to hold true, values of cloud
density and radius must be found so that for each experiment,
\beqa
F_{\rm c}&\geq &E_R F(E_R, 0)\int{ds\over M_Z^2}
\exp[-\sigma(s)n_\nu R]\nonumber\\
&\times&\biggl(1-\exp[\sigma(s)n_\nu R_c]\biggr).\label{coincident}
\eeqa
The AGASA experiment currently has the largest exposure at 
\beqa
{\cal
A}t_r = 670\ {\rm km^2\ sr\ year}\simeq 2\times 10^{21} {\rm cm^2\ sr\ 
s}\ .  
\eeqa
This allows a coincident flux of $F_{\rm c} \simeq 5
\times 10^{-22}\flux$. The values of $R$ and $n_\nu$ which solve
Eq.~\ref{coincident} are shown in Figs.~\ref{large}~and~\ref{small}. 
The coincidence constraint allows a minimum cloud radius of $R_{\rm
min} = 10^{18}{\rm cm}$ to be specified, and in some sense a maximum
relic neutrino density, depending on the incident neutrino flux.  The
results of this calculation, and the bounds from the UHECR data serve
to define an allowed parameter region for the neutrino cloud; see the
shaded area in Figs.~2~and~3.  We emphasize that, while the general 
principle that the non-observation of coincident events will eliminate 
a range of cloud radii less that a some ``minimum'' radius,  the 
particular 
value of this minimum radius is determined only after a  
model for the cloud, is chosen, and the  number density is calculated.  The 
uniform cloud considered here is but the simplest of the cloud 
models introduced in Ref.~(\cite{Mckellar1}), and even in that paper 
only a few simple cases were considered.  These other models, with 
 considerations similar to the above, will lead 
to different values for the minimum radius of the cloud.  We do not 
pursue this discussion into limitations regarding the variety of possible 
clouds in the present paper.

The constraints on the parameter range of neutrino clouds as a result
of the arguments based on coincident events have not taken into
account the spatial resolution of the various CR experiments.  While
this is not an issue for medium to large clouds, it does become
important for clouds with smaller radii.  For example, if a neutrino
cloud had a radius of $50{\rm au}$ decay products would be confined to
a cone of radius less than $10{\rm m}$ at earth.  This radius would be
too small for most CR experiments to resolve, in which case the decay
products will be measured as a single event with an energy of $E_R$.
In the minimal Z-burst model $E_R$ is a significantly higher energy than the
highest energy CR event.  However as we have shown in this note, for
high density clouds the effect of the Fermi motion is to reduce the
average value of $E_R$ so that the possibility of small radii clouds is
not excluded.  We leave a full quantitative analysis to future work,
and for the time being acknowledge that there is a region of allowed
parameter space for small radii neutrino clouds.

\section{Conclusion}
We have shown it is possible that neutrino clouds,
interacting with UHE neutrinos, are capable of producing the measured
flux of UHECR's, and that this result is within both theoretical
and experimental bounds imposed on the incident neutrino flux.  The
dimensions of the neutrino clouds required to produce the measured UHE
CR flux assuming the Z-burst mechanism is responsible were
investigated, and  need no new theoretical
treatment beyond that given in Ref.~(\cite{Mckellar1}).  The requirement
that no coincident events are observed requires that uniform clouds, larger
than some small radius still to be determined, are 
at least $0.3\rm{pc}$ in radius --- rather larger than originally 
envisaged in Ref.~(\cite{Mckellar1}), but without disturbing the 
viability of the concept of neutrino clouds.  Clouds with different 
density profiles will have different limitations placed on their radii.

\section*{Acknowledgments}
This research is supported in part by the Department of Energy under
contract W-7405-ENG-36, in part by the National Science Foundation 
and in part by the Australian Research Council.


\newpage
\begin{figure}
\begin{picture}(0,0)%
\includegraphics{fn-19.plot.pstex}%
\end{picture}%
\setlength{\unitlength}{3947sp}%
\begingroup\makeatletter\ifx\SetFigFont\undefined%
\gdef\SetFigFont#1#2#3#4#5{%
  \reset@font\fontsize{#1}{#2pt}%
  \fontfamily{#3}\fontseries{#4}\fontshape{#5}%
  \selectfont}%
\fi\endgroup%
\begin{picture}(5786,3495)(317,-3061)
\put(857,-2751){\makebox(0,0)[rb]{\smash{\SetFigFont{10}{12.0}{\familydefault}{\mddefault}{\updefault}10}}}
\put(857,-2366){\makebox(0,0)[rb]{\smash{\SetFigFont{10}{12.0}{\familydefault}{\mddefault}{\updefault}10.5}}}
\put(857,-1981){\makebox(0,0)[rb]{\smash{\SetFigFont{10}{12.0}{\familydefault}{\mddefault}{\updefault}11}}}
\put(857,-1596){\makebox(0,0)[rb]{\smash{\SetFigFont{10}{12.0}{\familydefault}{\mddefault}{\updefault}11.5}}}
\put(857,-1211){\makebox(0,0)[rb]{\smash{\SetFigFont{10}{12.0}{\familydefault}{\mddefault}{\updefault}12}}}
\put(857,-826){\makebox(0,0)[rb]{\smash{\SetFigFont{10}{12.0}{\familydefault}{\mddefault}{\updefault}12.5}}}
\put(857,-441){\makebox(0,0)[rb]{\smash{\SetFigFont{10}{12.0}{\familydefault}{\mddefault}{\updefault}13}}}
\put(857,-56){\makebox(0,0)[rb]{\smash{\SetFigFont{10}{12.0}{\familydefault}{\mddefault}{\updefault}13.5}}}
\put(857,329){\makebox(0,0)[rb]{\smash{\SetFigFont{10}{12.0}{\familydefault}{\mddefault}{\updefault}14}}}
\put(931,-2875){\makebox(0,0)[b]{\smash{\SetFigFont{10}{12.0}{\familydefault}{\mddefault}{\updefault}15}}}
\put(1447,-2875){\makebox(0,0)[b]{\smash{\SetFigFont{10}{12.0}{\familydefault}{\mddefault}{\updefault}15.5}}}
\put(1963,-2875){\makebox(0,0)[b]{\smash{\SetFigFont{10}{12.0}{\familydefault}{\mddefault}{\updefault}16}}}
\put(2479,-2875){\makebox(0,0)[b]{\smash{\SetFigFont{10}{12.0}{\familydefault}{\mddefault}{\updefault}16.5}}}
\put(2995,-2875){\makebox(0,0)[b]{\smash{\SetFigFont{10}{12.0}{\familydefault}{\mddefault}{\updefault}17}}}
\put(3511,-2875){\makebox(0,0)[b]{\smash{\SetFigFont{10}{12.0}{\familydefault}{\mddefault}{\updefault}17.5}}}
\put(4027,-2875){\makebox(0,0)[b]{\smash{\SetFigFont{10}{12.0}{\familydefault}{\mddefault}{\updefault}18}}}
\put(4543,-2875){\makebox(0,0)[b]{\smash{\SetFigFont{10}{12.0}{\familydefault}{\mddefault}{\updefault}18.5}}}
\put(5059,-2875){\makebox(0,0)[b]{\smash{\SetFigFont{10}{12.0}{\familydefault}{\mddefault}{\updefault}19}}}
\put(5575,-2875){\makebox(0,0)[b]{\smash{\SetFigFont{10}{12.0}{\familydefault}{\mddefault}{\updefault}19.5}}}
\put(6091,-2875){\makebox(0,0)[b]{\smash{\SetFigFont{10}{12.0}{\familydefault}{\mddefault}{\updefault}20}}}
\put(425,-1211){\rotatebox{90.0}{\makebox(0,0)[b]{\smash{\SetFigFont{10}{12.0}{\familydefault}{\mddefault}{\updefault}$\log(n_c)\ \rm{cm^{-3}}$}}}}
\put(3511,-3061){\makebox(0,0)[b]{\smash{\SetFigFont{10}{12.0}{\familydefault}{\mddefault}{\updefault}$\log(R)\ \rm{cm}$}}}
\end{picture}
\caption{The solid line represents the cloud parameters required to
produce a GZK flux of $F_{p/\gamma}=10^{-20}\flux$, with an incident
neutrino flux of $E_R F_\nu(E_R,0) = 10^{-19}\flux$.  The dashed line
represents a GZK flux of $F_{p/\gamma}=10^{-19}\flux$.  The short
dashed line is the bound on R resulting from the non-observation of
coincident events with an incident neutrino flux of $E_R F_\nu(E_R,0)
= 10^{-19}\flux$.  The shaded region is the allowed parameter space.}
\label{large}
\end{figure}
\begin{figure}
\begin{picture}(0,0)%
\includegraphics{fn-20.plot.pstex}%
\end{picture}%
\setlength{\unitlength}{3947sp}%
\begingroup\makeatletter\ifx\SetFigFont\undefined%
\gdef\SetFigFont#1#2#3#4#5{%
  \reset@font\fontsize{#1}{#2pt}%
  \fontfamily{#3}\fontseries{#4}\fontshape{#5}%
  \selectfont}%
\fi\endgroup%
\begin{picture}(5786,3495)(317,-3061)
\put(857,-2751){\makebox(0,0)[rb]{\smash{\SetFigFont{10}{12.0}{\familydefault}{\mddefault}{\updefault}10}}}
\put(857,-2366){\makebox(0,0)[rb]{\smash{\SetFigFont{10}{12.0}{\familydefault}{\mddefault}{\updefault}10.5}}}
\put(857,-1981){\makebox(0,0)[rb]{\smash{\SetFigFont{10}{12.0}{\familydefault}{\mddefault}{\updefault}11}}}
\put(857,-1596){\makebox(0,0)[rb]{\smash{\SetFigFont{10}{12.0}{\familydefault}{\mddefault}{\updefault}11.5}}}
\put(857,-1211){\makebox(0,0)[rb]{\smash{\SetFigFont{10}{12.0}{\familydefault}{\mddefault}{\updefault}12}}}
\put(857,-826){\makebox(0,0)[rb]{\smash{\SetFigFont{10}{12.0}{\familydefault}{\mddefault}{\updefault}12.5}}}
\put(857,-441){\makebox(0,0)[rb]{\smash{\SetFigFont{10}{12.0}{\familydefault}{\mddefault}{\updefault}13}}}
\put(857,-56){\makebox(0,0)[rb]{\smash{\SetFigFont{10}{12.0}{\familydefault}{\mddefault}{\updefault}13.5}}}
\put(857,329){\makebox(0,0)[rb]{\smash{\SetFigFont{10}{12.0}{\familydefault}{\mddefault}{\updefault}14}}}
\put(931,-2875){\makebox(0,0)[b]{\smash{\SetFigFont{10}{12.0}{\familydefault}{\mddefault}{\updefault}16}}}
\put(1576,-2875){\makebox(0,0)[b]{\smash{\SetFigFont{10}{12.0}{\familydefault}{\mddefault}{\updefault}16.5}}}
\put(2221,-2875){\makebox(0,0)[b]{\smash{\SetFigFont{10}{12.0}{\familydefault}{\mddefault}{\updefault}17}}}
\put(2866,-2875){\makebox(0,0)[b]{\smash{\SetFigFont{10}{12.0}{\familydefault}{\mddefault}{\updefault}17.5}}}
\put(3511,-2875){\makebox(0,0)[b]{\smash{\SetFigFont{10}{12.0}{\familydefault}{\mddefault}{\updefault}18}}}
\put(4156,-2875){\makebox(0,0)[b]{\smash{\SetFigFont{10}{12.0}{\familydefault}{\mddefault}{\updefault}18.5}}}
\put(4801,-2875){\makebox(0,0)[b]{\smash{\SetFigFont{10}{12.0}{\familydefault}{\mddefault}{\updefault}19}}}
\put(5446,-2875){\makebox(0,0)[b]{\smash{\SetFigFont{10}{12.0}{\familydefault}{\mddefault}{\updefault}19.5}}}
\put(6091,-2875){\makebox(0,0)[b]{\smash{\SetFigFont{10}{12.0}{\familydefault}{\mddefault}{\updefault}20}}}
\put(425,-1211){\rotatebox{90.0}{\makebox(0,0)[b]{\smash{\SetFigFont{10}{12.0}{\familydefault}{\mddefault}{\updefault}$\log(n_c)\ \rm{cm^{-3}}$}}}}
\put(3511,-3061){\makebox(0,0)[b]{\smash{\SetFigFont{10}{12.0}{\familydefault}{\mddefault}{\updefault}$\log(R)\ \rm{cm} $}}}
\end{picture}
\caption{The solid line represents the cloud parameters required to
produce a GZK flux of $F_{p/\gamma}=10^{-20}\flux$, with an incident
neutrino flux of $E_R F_\nu(E_R,0) = 10^{-20}\flux$.  The dashed line
represents a GZK flux of $F_{p/\gamma}=1.7\times 10^{-20}\flux$, the
maximum flux for this small incident neutrino flux.  The short
dashed line is the bound on R resulting from the non-observation of
coincident events with an incident neutrino flux of $E_R F_\nu(E_R,0)
= 10^{-20}\flux$.  The shaded region is the allowed parameter space. }
\label{small}
\end{figure}
%
%

\end{document}